\documentstyle[epsf,twoside,fleqn,espcrc2]{article}

\newcommand{\wone}{{$\Box_{1\times 1}$}}
\newcommand{\weight}{{$\Box_{8\times 8}$}}
\newcommand{\wsixt}{{$\Box_{16\times 16}$}}

\newcommand{\tkl}{{{\small\sf T}{$\chi$}{\small\sf L}}}
\newcommand{\sesam}{{\small\sf SESAM}}

\newcommand{\bi}{\begin{itemize}}
\newcommand{\ei}{\end{itemize}}

\newcommand{\beq}{\begin{equation}}
\newcommand{\eeq}{\end{equation}}

\newcommand{\fig}[1]{Fig.~\ref{#1}}
\newcommand{\tab}[1]{Tab.~\ref{#1}}

\newcommand{\ks}{\kappa_{\rm sea}}

\newcommand{\kc}{\kappa_{\rm sea}^c}

\newcommand{\prd}[3]{ \frenchspacing Phys.      Rev.  D {\bf #1}
  {(#2)} {#3}} 
\newcommand{\npb}[3]{ \frenchspacing Nucl. Phys.  B {\bf #1}
  {(#2)} {#3}} 
\newcommand{\plb}[3]{ \frenchspacing Phys.      Lett. B {\bf #1}
  {(#2)} {#3}}

\newcommand{\ie}{{\frenchspacing\em i.\hspace{0.4mm}e.{}}}

\newcommand{\AmS}{{\protect\the\textfont2
  A\kern-.1667em\lower.5ex\hbox{M}\kern-.125emS}}

\title{ Critical Dynamics of the Hybrid Monte Carlo Algorithm}
     
\author{\frenchspacing \sesam\ $+$ \tkl-Collaboration: Th.
  Lippert$^{\rm{b}}$\thanks{Presented by Th.  Lippert}, G.
  Bali\address{Physics Department, Humboldt University, Berlin,
    Germany\vspace{-.2cm}}, N. Eicker\address{HLRZ, c/o J\"ulich
    Research Center and DESY, Hamburg, D-52425-J\"ulich,
    Germany\vspace{-.2cm}}, L. Giusti\address{Scuola Normale
    Superiore, I-56100 Pisa and INFN Sez. di Pisa,
    Italy\vspace{-.2cm}}, U.  Gl\"assner\address{Department of
    Physics, University of Wuppertal, D-42097 Wuppertal,
    Germany\vspace{-.2cm}}, S.  G\"usken$^{\rm{d}}$, H.
  Hoeber$^{\rm{d}}$, G. Martinelli\address{INFN, University ``La
    Sapienza'', P'lle Aldo Moro, Roma, Italy\vspace{-,2cm}}, F.
  Rapuano$^{\rm{e}}$, G.  Ritzenh\"ofer$^{\rm{b}}$, K.
  Schilling$^{\rm{b,d}}$, A.  Spitz$^{\rm{d}}$, and J.
  Viehoff$^{\rm{d}}$\nonfrenchspacing}

\begin{document}

\begin{abstract}
  We investigate the critical dynamics of the Hybrid Monte Carlo
  algorithm approaching the chiral limit of standard Wilson fermions.
  Our observations are based on time series of lengths $O(5000)$ for a
  variety of observables.  The lattice sizes are $16^3\times 32$ and
  $24^3\times 40$. We work at $\beta=5.6$, and $\kappa=0.156$,
  $0.157$, $0.1575$, $0.158$, with $0.83 > \frac{m_{\pi}}{m_{\rho}} >
  0.55$.  We find surprisingly small integrated autocorrelation times
  for local and extended observables.  The dynamical critical exponent
  $z$ of the exponential autocorrelation time is compatible with 2.
  We estimate the total computational effort to scale between $V^{2}$
  and $V^{2\frac{1}{4}}$ towards the chiral limit.
\end{abstract}
\maketitle

\section{INTRODUCTION}
Considerable effort has been spent since the early days of exact full
QCD simulations with Hybrid Monte Carlo methods \cite{DUANE} to
optimize both performance and de-correlation efficiency. The main
issues are {\em parameter tuning}, {\em algorithmic improvements} and
{\em assessment of de-correlation}.

HMC parameter tuning has been the focus of the early papers.  At this
time one was restricted to quite small lattices of size $\approx 4^4$
\cite{CREUTZ89,GUPTATUN89,BITAR87}.  As main outcome, trajectory
lengths of $O(1)$ and acceptance rates of $\approx70$ \% have been
recommended.

The last three years have seen a series of improvements: Inversion
times could be reduced by use of the BiCGstab algorithm
\cite{FROMMER94,DEFORCRAND94} and parallel SSOR preconditioning
\cite{FISCHER96}.  Together with chronological inversion
\cite{BROWER96}, gain factors between 4 and 8 could be realized
\cite{TSUKUBA}.

Less research has been done on the critical dynamics of HMC.  In
Ref.~\cite{PETERSEN}, the computer time required to de-correlate a
staggered fermion lattice has been estimated to grow like $T\propto
m_q^{\frac{-21}{4}}$. A similar law has been guessed in
Ref.~\cite{GUPTA91} with $T\propto m_{\pi}^{\frac{-23}{2}}$ for two
flavours of Wilson fermions.  However, so far we lack knowledge about
the scaling of the critical behaviour of HMC: a clean determination of
the autocorrelation times requires by far more trajectories than
feasible at the time.  Ref.~\cite{JANSENLIU} quotes results in SU(2),
the APE group found autocorrelation times in the range of 50 for meson
masses \cite{APE}, the SCRI group gives estimates for $\tau_{int}$
between 9 and 65. However, they seem to behave inconsistently with
varying quark mass \cite{SCRI95}.

\sesam\ and \tkl\ have boosted the trajectory samples to $O(5000)$,
\ie, by nearly one order of magnitude compared to previous studies
\cite{TSUKUBA}.  The contiguous trajectories are generated under
stable conditions and allow for reliable determinations of
autocorrelation times from HMC simulations.

\section{AUTOCORRELATION TIMES}

The finite time-series approximation to the true autocorrelation
function for $A_t$, $t=1,\dots,t_{MC}$ is
\begin{eqnarray}
C^A(t)= \frac{\sum\limits_{s=1}^{t_{MC}-t} A_sA_{s+t} - 
\frac{1}{t_{MC}-t}
\left(\sum\limits_{s=1}^{t_{MC}}
A_s \right) ^2}{t_{MC}-t}.
\end{eqnarray}
The {\em exponential} autocorrelation time is the inverse decay rate
of the slowest contributing mode with $\rho^A(t)$ being $C^A(t)$
normalized to $\rho^A(0)=1$, $ \tau^A_{exp} =
\limsup_{t\rightarrow\infty}\frac{t}{-\log\rho^A(t)} $.
$\tau^A_{exp}$ is related to the length of the thermalization phase of
the Markov process.  To achieve ergodicity the simulation has to
safely exceed $\mbox{sup}_A \{\tau^A_{exp}\}$.  The {\em integrated}
autocorrelation time reads:
\begin{equation}
\tau^A_{int}=\frac{1}{2}+\sum_{t'=1}^{t_{MC} \to \infty}\rho^A(t').
\label{tauint}
\end{equation}
In equilibrium $\tau^A_{int}$ characterizes the true statistical error
of the observable $A$. The variance $\sigma_A^2 =
2\tau^A_{int}\sigma^2_{0}$ is increased by the factor $2\tau^A_{int}$
compared to the result over a sample of $N$ independent
configurations.  $\tau^A_{int}$ is observable dependent.

\section{RESULTS}
The results given here can only be a small excerpt of our
investigation to be presented elsewhere in more detail \cite{LIPPERT}.
The autocorrelation is determined from the plaquette (\wone) and
extended quantities like smeared light meson masses, $m_\pi$ and
$m_\rho$, and smeared spatial Wilson loops (\weight\ and \wsixt).
They exhibit a large ground state overlap per construction.
Furthermore we exploited the inverse of the average number of
iterations $\Lambda=N_{kry}^{-1}$ of the Krylov solver which is
related to the square root of the ratio of the minimal to the maximal
eigenvalue of the fermion matrix \cite{TSUKUBA}.

We illustrate the observable dependency of the autocorrelation in
\fig{OBS}.
\begin{figure}[tb]
\centerline{
\epsfxsize=.4\textwidth\epsfbox{obs.eps}}
\vspace*{-1cm}
\caption{
  $\rho(t)$ at $\ks=0.157$ for \wone, and $\Lambda$.}
\label{OBS} 
\end{figure}
Using $O(3000)$ trajectories we achieve a clear signal for the
autocorrelation function. This is the case also for the dynamical
samples at $\ks=0.156$, $0.157$ and $0.1575$ as well as on the
$24^3\times 40$ lattice with $\ks=0.1575$ and $0.158$. $\Lambda$
appears to give an upper estimate to the autocorrelation times of
`fermionic' and `gluonic' observables measured\footnote{Slower modes
  might exist for the topological charge \cite{BOYD}.}.  

\begin{figure}
\centerline{
\epsfxsize=.4\textwidth\epsfbox{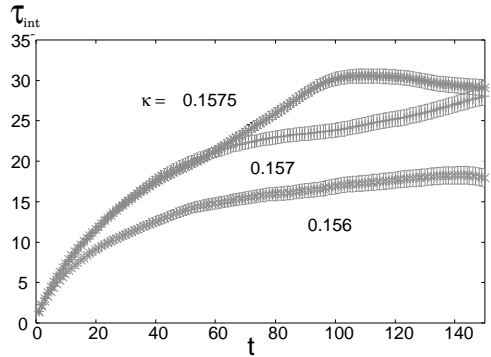}}
\vspace*{-1cm}
\caption{
  $\ks$ dependency of $\Lambda$ on the $16^3\times 32$ lattice.}
\label{SEA} 
\end{figure}

Shifting the sea-quark mass towards the chiral limit, we observe
increasing $\tau_{exp}$ and $\tau_{int}$. \fig{SEA} sketches this
dependence of $\tau_{int}^{\Lambda}$ on the $16^3\times 32$ lattice.

The volume dependence of the autocorrelation was compared for two
different lattice sizes at equal $\ks=0.1575$. Unexpectedly, we found
$\tau_{exp}$ and $\tau_{int}$, measured in units of trajectory
numbers, to decrease by about 50 \% for \wone\ and 30 \% for $\Lambda$,
while switchng from the $16^3$ to the $24^3$ system. As we chose the
length of the HMC molecular dynamics as $T=1$ on the $16^3\times 32$
system and $T=0.5$ on the $24^3\times 40$ lattice, these numbers are
even more surprising.  The origin of this phenomenon needs further
investigations.

\begin{figure}[!htb]
\centerline{
\epsfxsize=.4\textwidth\epsfbox{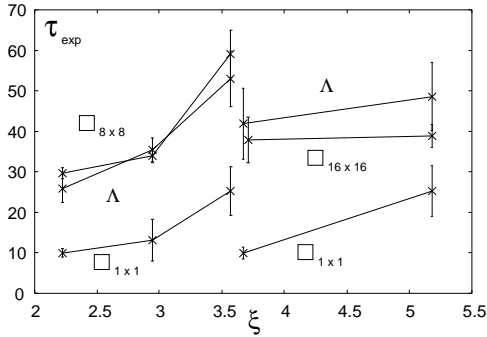}}
\vspace*{-1cm}
\caption{
Exponential autocorrelation times.}
\label{COMPILATION} 
\end{figure}

A compilation of $\tau_{exp}$ is given in \fig{COMPILATION}, the
values for $\tau_{int}$ being similar.

The quality of our data allows to address the issue of critical
slowing down for HMC.  The approach to $\kc$ amounts to a growing pion
correlation length $\xi_{\pi}=1/m_{\pi}a$.  The autocorrelation time
is expected to scale with a power of $\xi$, $\tau=\epsilon\xi^z$.  The
{\em dynamical critical exponent} $z$ governs the scaling of the
compute effort, while the knowledge of $\epsilon$ allows to assess its
absolute value. Our results are given in \tab{RES}.
\begin{table*}[!htb]
\setlength{\tabcolsep}{1.5pc}
\newlength{\digitwidth} \settowidth{\digitwidth}{\rm 0}
\catcode`?=\active \def?{\kern\digitwidth}
\caption{Critical exponents $z$. The $16^3\times 32$ results are
  based on full, the $24^3\times 40 $
numbers on partial statistics.}
\label{RES}
\begin{tabular*}{\textwidth}{@{}l|@{\extracolsep{\fill}}ccc | ccc}
\hline
Size          & \multicolumn{3}{c|}{$16^3\times 32$} &\multicolumn{3}{c}{$24^3\times 40$}\\
\hline
Observable    &  \wone\  & $\Lambda$ & \weight\  &  \wone\  & $\Lambda$ & \wsixt \\
\hline
$z_{exp}$     &  1.8(4) & 1.5(4) & 1.2(5) & 2.2 & 0.5 & 0.1 \\
\hline
$z_{int}$     &  1.4(7) & 1.3(3) & 1.3(3) & 2.6 & 1.0 & 0.3 \\
\hline
\end{tabular*}
\end{table*}
For the smaller lattice they tend to be below $z=2$, albeit they are
still compatible with 2.  It seems that extended observables on the
large lattice yield smaller values. However, we have to wait for
larger samples to arrive at conclusive answers.

Finally we try to give a conservative guess for the scaling of the
compute time $T$ required for de-correlation.  The maximal length of
$\xi_{\pi}$ has been limited to $V^{\frac{1}{4}}/\xi_{\pi}\approx 4$
in order to avoid finite size effects. With $\xi_{\pi}$ fixed, the
volume factor goes as $\xi_{\pi}^4$. Furthermore we found
$\Lambda\propto\xi^{-2.7}$ for BiCGstab. $z$ lies between 1.3 and 1.8,
taking the result from the $16^3$ system.

In order to keep the acceptance rate constant, we reduced the time
step from 0.01 to 0.004 with increasing lattice size, while we have
increased the number of HMC time steps from 100 to 125.

At the same time, as we have already mentioned, the autocorrelation
time for the worst case observable $\Lambda$ went down by 30 \%
compensating the increase in acceptance rate cost!  In a conservative
estimate, we thus would assume that the total time $T$ scales as $
\xi_{\pi}^8$ to $\xi_{\pi}^{8.5}$. This guess translates to $T\propto
V^2$ - $V^{2.2}$.

\section{SUMMARY}

The autocorrelation times from HMC under realistic conditions are
smaller than anticipated previously, staying below the value of $60$
trajectories for $\tau_{exp}$.  $z$ is compatible with 2.  The
computer time to generate decorrelated configurations scales according
to $V^2$, a very encouraging result compared to previous estimates
\cite{PETERSEN,GUPTA91}.

\end{document}